\documentclass[twocolumn,pra,superscriptaddress,showpacs]{revtex4}

\usepackage{amsmath, upgreek, amssymb}
\usepackage{graphicx, subfigure}

\newcommand{\eref}[1]{Eq.~(\ref{#1})}
\newcommand{\Eref}[1]{Equation~(\ref{#1})}
\newcommand{\fref}[1]{Fig.~\ref{#1}}
\newcommand{\Fref}[1]{Fig.~\ref{#1}}
\newcommand{\sref}[1]{section~\ref{#1}}
\newcommand{\Sref}[1]{Section~\ref{#1}}
\newcommand{\heatingcoefft}{\Upsilon}
\newcommand{\ie}{\emph{i.e.}}
\newcommand{\rmd}{\text{d}}
\newcommand{\force}{\boldsymbol{F}}
\newcommand{\diffn}{\boldsymbol{D}}
\DeclareMathOperator{\sinc}{sinc}

\makeatletter
\newlength \figwidth
\makeatother

\begin{document}

\def \initials {AX}

\title{Atom cooling using the dipole force of a single retroflected laser beam}
\author{Andr\'e Xuereb}
\email[To whom all correspondence should be addressed. Electronic address:\
]{andre.xuereb@soton.ac.uk}
\affiliation{School of Physics and Astronomy, University of Southampton, Southampton SO17~1BJ, United Kingdom}
\author{Peter Horak}
\affiliation{Optoelectronics Research Centre, University of Southampton, Southampton SO17~1BJ, United Kingdom}
\author{Tim Freegarde}
\affiliation{School of Physics and Astronomy, University of Southampton, Southampton SO17~1BJ, United Kingdom}

\date{\today}


\begin{abstract}
We present a mechanism for cooling atoms by a laser beam reflected from a
single mirror. The cooling relies on the dipole force and thus in principle
applies to arbitrary refractive particles including atoms, molecules, or
dielectric spheres. Friction and equilibrium temperatures are derived by an
analytic perturbative approach. Finally, semiclassical Monte-Carlo simulations
are performed to validate the analytic results.
\end{abstract}
\maketitle

\section{Introduction}

Optical cooling of atoms has come a long way since it was first
proposed; magneto--optical traps are even found in undergraduate
laboratories~\cite{Hansch1975, Wineland1979, Mellish2002}. The field of
ultracold molecules,
in contrast, is still in its infancy. Ultracold diatomic alkali
molecules ($<100$~$\upmu$K for Rb$_2$~\cite{Gabbanini2000}) are
routinely produced from Bose-Einstein condensates through Feshbach
resonances. Some groups (see, for example,~\cite{Sage2005}
and~\cite{Winkler2007}) have demonstrated the possibility of cooling
the internal degrees of such molecules, cooling ultracold di-alkali
molecules to their lowest rovibronic levels by means of
laser-stimulated state transfer processes.

Present methods of producing ultracold samples suffer from one of
two major drawbacks: either they are specific to particular species,
or they produce very low densities. The bulk of optical cooling
methods are applicable only to a handful of species because they
rely on a closed optical transition within which the population can
cycle~\cite{Metcalf2003}. Most atoms and molecules do not have such
a transition available, but instead exhibit a large number of loss
channels through which the population is gradually lost, halting the
cooling. Samples of cold molecules are therefore generally produced
by capturing the low-velocity tail of the Maxwell-Boltzmann
distribution of a hotter initial sample \cite{Pinkse2003}. However,
such filtering methods do not lead to an increase of the phase-space
density and thus only capture a small fraction of the initial
population, leading to very dilute samples.

One possibility to solve this problem is through the use of
non-resonant processes \cite{Kerman2000} or cavities
\cite{Horak1997, Vuletic2000, Maunz2004, Vilensky2007, Lev2008}. The
latter require extremely high precision alignment of the cavity
mirrors as well as complicated loading of the molecules into the
optical cavity mode. The requirements for integrated systems near
the surface of a substrate in the form of atom chips
\cite{Folman2002} are even more stringent.

Here, we investigate a mechanism for the cooling of a particle
using only a single plane mirror in place of a cavity. In principle,
this scheme only relies on the dipole force of a refractive particle
in a laser beam and thus applies to a wide range of atomic and
molecular species as well as, for example, dielectric micro- or
nanospheres. However, in the present work we focus on the basic
principles of the cooling scheme and thus restrict the analysis
to the simplest case of a two-level atom.

Conceptually, one can view the interaction between the atom and the mirror as being closely related to that between a micromechanical oscillator, acting as a mobile mirror, and a second mirror. Such schemes have been investigated both theoretically \cite{Braginsky2002, Bhattacharya2008} and experimentally \cite{Corbitt2007, Schliesser2008} in various configurations.\\
Although we have recently shown that one can treat these two situations as two opposite limits of the same model \cite{Xuereb2009b}, the situation we explore here behaves differently, and this can be attributed to two facts. Firstly, the coupling strength between the static and moving scatterer (atom or mirror) is very different in the two cases: an atom merely perturbs the field it interacts with, whereas a mirror acts as a moving boundary condition and changes the field significantly. Secondly, the effect we investigate here is only dominant at large atom--mirror separations. Thus, our proposed cooling scheme operates in a parameter regime that is as yet mostly unexplored.

This paper is structured as follows. In the next section we
introduce the key features of our system and propose a simple
classical explanation of the cooling scheme. In \sref{sec:System}
the relevant quantum equations of motion are introduced.
\Sref{sec:Perturbative} solves these equations of motion
analytically through the use of perturbative methods, whereas
\sref{sec:Numerical} gives the results of numerical simulations used
to explore the implications of the theory in further detail.
\Sref{sec:Beyond} compares our results with those of traditional
Doppler cooling, and finally \sref{sec:Conclusion} summarizes and
concludes our discussion.

\section{Motivation}
\label{sec:Motivation}

\begin{figure}[t]
 \centering
    \includegraphics[width=\figwidth]{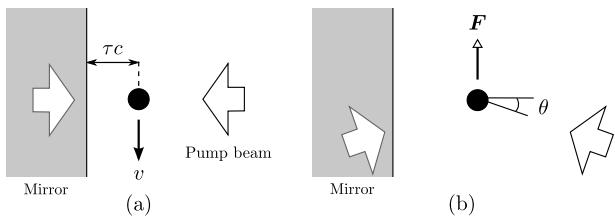}
\caption{Schematic of the cooling scheme. (a) In the laboratory
frame, an atom moving with velocity $v$ parallel to a mirror, a
distance $\tau c$ away, interacts with a pump beam and its
time-delayed reflection. (b) In the frame of the atom, the
(relativistically transformed) pump beam and its reflection are
tilted by an angle $\theta=\arcsin(v/c)$ and produce a net retarding
force, $\force$.}
 \label{fig:Motivation}
\end{figure}

We start with a classical explanation of the situation, which provides the
motivation for the mathematical model presented in the
next section. The phenomenon of optical binding has been known for
some time (see, for example, \cite{Burns1990, MacDonald2002,
Metzger2006a}) and is now a common occurrence. At a basic level,
optical binding takes place between two dielectric spheres when one
sphere focuses the light onto a second sphere, which is
subsequently trapped. If we now consider just one such sphere in
front of a mirror, the modified electric field will be reflected
back towards the sphere itself. In essence, then, the sphere will be
attracted to its own image. However, this interaction is delayed by
the time $2\tau$ it takes the disturbance in the electric field to
travel from the sphere to the mirror and back, where $\tau$ is the time that
light from the atom takes to reach the mirror. Suppose, now, that
the sphere is moving parallel to the plane of the mirror with
velocity $v$, as shown in \Fref{fig:Motivation}(a). In this case,
the sphere moves a distance $2v\tau$ during the light roundtrip
time. Thus, the disturbed light field lags behind the particle and
creates an attractive force in the direction opposite to the motion
of the particle. This attractive force can be shown to be a viscous
force, \ie, it is proportional to $v$. This scenario also applies to
the case of a single atom interacting with an off-resonant beam,
where the atom can effectively be modeled as a refractive particle.
Note that the interaction between a single atom and its image in a
distant mirror has already been demonstrated \cite{Eschner2001,Bushev2004}.

Alternatively, we may consider the same situation in the reference
frame of the moving particle, as shown in \fref{fig:Motivation}(b).
In this case both the incident laser beam and its reflection are
tilted by an angle $\theta=\arcsin(v/c)$ with respect to normal
incidence on the mirror. The sum potential is therefore offset from
the particle position by an amount proportional to its velocity,
leading again to a velocity-dependent force opposing the motion.

Similar arguments apply for a particle moving along the direction of
the pump beam, \ie, orthogonal to the mirror. In this case, the
phase of the reflected beam is determined by the interaction of the
particle with the pump at the earlier time $2\tau$. In effect, the
atom exerts a delayed phase change on the light field, dragging the
potential along with itself while moving and thus creating a
non-conservative force. This effect is similar to the ``position
dependent phase locking'' described in \cite{Gangl2000a}.

\section{Mathematical model\label{sec:System}}
\begin{figure}[t]
 \centering
    \includegraphics[scale=0.6]{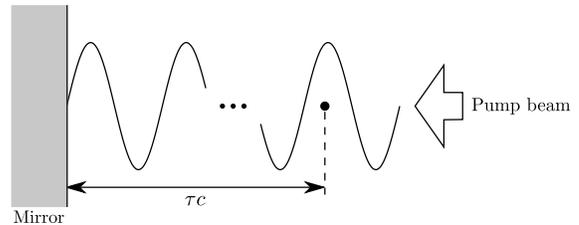}
 \caption{Schematic representation of the key components of the
 system under consideration. The atom is separated from the mirror
 by a distance $\tau c$ and lies in a standing wave maintained by
 the pump beam.}
 \label{fig:Model}
\end{figure}

In order to analyze the principles of the proposed cooling scheme
most clearly, we simplify the situation described above to a
one-dimensional (1D) scheme and assume a single two-level atom as
the particle to be cooled. A schematic of the system is shown in
\Fref{fig:Model}.

The atom has a transition frequency $\omega_{\text{a}}$ and a decay
rate $2\Gamma$ and is described by the operators $\hat{p}$ and
$\hat{x}$ associated with the atomic momentum and position,
respectively, and by the atomic dipole raising $(\hat{\sigma}^+)$
and lowering $(\hat{\sigma}^-)$ operators. The atom is coupled to a
continuum of quantized electro-magnetic modes with frequencies
$\omega$ and standing-wave mode functions $f(\omega,x) = \sin(\omega
x/c)$, described by the field annihilation $\hat{a}(\omega)$ and
creation operators $\hat{a}^{\dagger}(\omega)$. The mirror is at
position $x=0$. For simplicity, we neglect the frequency-dependence
of the atom-field coupling and assume a single coupling coefficient
$g$. Finally, mode $\omega_0$ is pumped by a laser, which enters our
analysis as an initial condition, and far off resonant pumping is
assumed, $|\Delta|=|\omega_a-\omega_0| \gg \Gamma$, where the atom
mainly acts as a refractive particle and spontaneous scattering is
reduced. For the numerical examples given in this paper, we consider
$^{85}$Rb atoms and a realistic pump beam that is detuned from the
$5{\rm S}_{1/2}\rightarrow5{\rm P}_{3/2}$ transition of $^{85}$Rb by
several linewidths.

The starting point for describing the coupling between the atom and
the field modes is the quantum master equation:
\begin{equation}
 \dot{\hat{\rho}}=-\frac{i}{\hbar}\left[\hat{H},\hat{\rho}\right]+\mathcal{L}
\hat{\rho},
 \label{eq:master}
\end{equation}
where $\hat{\rho}$ is the density operator of the full system
comprising all modes and the atom. Applying the dipole and rotating
wave approximations and in a frame rotating with the driving
frequency $\omega_0$, the Hamiltonian $\hat{H}$ reads
\begin{align}
\label{eq:Hamiltonian1}
 \hat{H}=&\,\frac{\hat{p}^2}{2m}
 +\hbar\Delta\hat{\sigma}^+\hat{\sigma}^-\!\!\int\hbar
(\omega-\omega_0)\hat{a}^{\dagger}(\omega)\hat{a}(\omega)\,\rmd\omega\nonumber\\
&-\ i\hbar g\!\!\int[\hat{\sigma}^+\hat{a}(\omega)f(\omega,\hat{x}) -
{\rm h.c.}]\,\rmd\omega\text{,}
\end{align}
and the damping term associated with atomic decay into modes other
than the 1D system modes reads
\begin{align}
\label{eq:Damping1}
\mathcal{L}\hat{\rho}=-\Gamma\bigg[&\hat{\sigma}^+\hat{\sigma}^-\hat{\rho} +
\hat{\rho}\hat{\sigma}^+\hat{\sigma}^-\nonumber\\
&-\
2\int_{-1}^{1}N(u)\hat{\sigma}^-\text{e}^{-iu\hat{x}}\hat{\rho}\text{e}^{iu\hat{
x}}\hat{\sigma}^+\,\rmd u\bigg]\text{.}
\end{align}
Here $N(u)$ describes the 1D projection of the spontaneous emission
pattern of the atomic dipole. In the low saturation regime we can
adiabatically eliminate the internal atomic dynamics and formally
express the dipole operator as
\begin{equation}
\label{eq:AdiabaticApprox}
 \hat{\sigma}^-=-\frac{i\Delta+\Gamma}{\Delta^2+\Gamma^2}\,g\int
f(\omega,\hat{x})\hat{a}(\omega)\,\rmd\omega+\hat{\xi}^-\text{,}
\end{equation}
where $\hat{\xi}^-$ is a noise term \cite{Gardiner1984}.

In the following, we present two different ways of proceeding
from this point. In the first instance we approximate further
and use perturbation theory to derive the force experienced by the
atom analytically, in \sref{sec:Perturbative}, and in the second
instance we derive semiclassical equations of motion. The
latter approach is then applied to numerical simulations of the
situation in \sref{sec:Numerical}.

\section{Analyzing the model: A perturbative approach}
\label{sec:Perturbative}

\subsection{Friction force}
\label{sec:PerturbativeFriction}

We first derive an analytical approximation for the friction on
the atom. To this end, we treat atomic motion semiclassically and
thus replace the operator $\hat{x}$ by an atomic position $x$. After
inserting \eref{eq:AdiabaticApprox} into \eref{eq:Hamiltonian1} and
\eref{eq:Damping1}, we obtain the Hamiltonian
 \begin{multline*}
\hat{H}=\int\hbar(\omega-\omega_0)\hat{a}^{\dagger}(\omega)\hat{a}(\omega)\,
\rmd\omega+\hbar
g^2D(\Delta)\\\times\iint\sin(\omega_1 x/c)\sin(\omega_2
x/c)\hat{a}^{\dagger}(\omega_1)\hat{a}(\omega_2)\,\rmd\omega_1\,
\rmd\omega_2\text{,}
 \end{multline*}
where we have defined $D(\Delta)=\Delta/(\Delta^2+\Gamma^2)$. As a
consequence of the assumed large pump detuning $\Delta$, we in
the following neglect that part of the decay term $\mathcal{L}\hat{\rho}$
which leads to spontaneous scattering of photons between the
quantized modes by the atom.

Let us first consider a stationary atom at a fixed position $x=x_0$.
Starting with the Heisenberg equation of motion for the annihilation
operators
\begin{equation*}
 \frac{\rmd}{\rmd
t}\hat{a}(\omega,t)=\frac{i}{\hbar}\left[\hat{H},\hat{a}(\omega,t)\right]\text{,
}
\end{equation*}
we arrive at the integro-differential equation
\begin{align}
\label{eq:StartingDE}
\frac{\rmd}{\rmd
t}\hat{a}(\omega,t)=&-i(\omega-\omega_0)\hat{a}(\omega,t)\nonumber\\
-\ i g^2 D(\Delta)&\sin(\omega x/c)\int\sin(\omega_1
x/c)\hat{a}(\omega_1,t)\,\rmd\omega_1\text{.}
\end{align}
We now assume coherent states at all times for the fields and
replace the operators with their respective expectation values.
Since we are pumping the atom at a single frequency, we take the
initial condition $a(\omega,0)=A\delta(\omega-\omega_0)$, where $A$
is the amplitude of the pump field, such that $|A|^2$ is the pump
power in units of photons per second, and $\delta$ is the Dirac
$\delta$-function. We now expand the fields $a(\omega,t)$ in the
weak-coupling limit in powers of the coupling constant,
\begin{equation}
a(\omega,t) = \sum_n a_n(\omega,t)[g^2 D(\Delta)]^n,
\end{equation}
with $a_n(\omega,t)$ being the $n$th coefficient of the series expansion.
Solving \eref{eq:StartingDE} by perturbation theory then
yields the zeroth order term in $g^2 D(\Delta)$
\begin{align}
\label{eq:a_0}
 a_0(\omega,t)&=A\delta(\omega-\omega_0)\text{,}
\end{align}
and the first order term
\begin{align*}
 a_1(\omega,t)=A\frac{\exp\left[-i(\omega-\omega_0)t\right]-1}{\omega-\omega_0}
\sin(\omega x/c)\sin(\omega_0 x/c)\text{.}
\end{align*}
We now proceed to find, to second order in $g^2 D(\Delta)$, the
static force, $\force(x_0,t)=-\partial\hat{H}/\partial x$, acting on
the atom:
 \begin{multline*}
  \force(x_0,t)=\frac{\hbar}{c}\left|A\right|^2 g^2
D(\Delta)\omega_0\bigg\{\sin(2\omega_0 x_0/c)\\-\frac{\pi}{2}g^2
D(\Delta)\sin^2(\omega_0 x_0/c)\big[4\cos^2(\omega_0 x_0/c)-1
\big]\bigg\}\text{.}
 \end{multline*}
The first term in the above equation describes the interaction
of the atom with the unperturbed pump field, whereas the second term
is the lowest order correction of the force due to the back-action
of the atom on the light fields. Note that this force is independent
of time.

Similarly, we can now calculate the force on an atom moving at a
constant velocity $v$. For this we assume that the atom follows a
trajectory given by $x(t)=x_0+v(t-t_0)$, where $t_0$ is a long
enough time for the system to reach a stationary state, \ie, $t_0$
is larger than twice the propagation time $\tau=x_0/c$ of the light
from the atom to the mirror. We can then solve \eref{eq:StartingDE} up to first
order in both $v$ and $g^2
D(\Delta)$. The friction force in the longitudinal direction is
finally obtained as
\begin{eqnarray}
\label{eq:AnalyticLongFrictionAll}
 \force_{\parallel}(x_0,t) &=& \frac{2\pi\hbar\omega_0}{c^2}v|A|^2\big[g^2
D(\Delta)\big]^2\sin^2(2k_0 x_0)
\nonumber \\
& &-\frac{2\pi\hbar\omega_0^2}{c^2}v\tau|A|^2\big[g^2
D(\Delta)\big]^2\sin(4k_0 x_0).
\end{eqnarray}
The second term in~\eref{eq:AnalyticLongFrictionAll} is larger than
the first by a factor of the order $k_0x_0=\omega_0 x_0/c$ and is
therefore dominant if the distance of the atom from the mirror is
much larger than an optical wavelength. We may then approximate the
longitudinal friction force by
\begin{equation}
\label{eq:AnalyticLongFriction}
 \force_{\parallel}(x_0,t)=-2\pi\hbar
k_0^2v\tau|A|^2\left(\frac{g^2\Delta}{\Delta^2+\Gamma^2}\right)^{\!\!2}
\sin(4k_0x_0)\text{.}
\end{equation}
Supposing that the species we are cooling is $^{85}$Rb, and setting
$|A|^2=62.5 \Gamma/(2\pi)$, $\Delta=-10\Gamma$,
$\tau=0.25/\Gamma$, and Gaussian beam waist
$w=0.7~\upmu$m,~\eref{eq:AnalyticLongFriction} predicts $1/e$
cooling times of the order of $2$~ms. The value for $\tau$ that we use implies a
separation between the atom and the mirror of the order of several metres. We
suggest that this problem can be overcome through the coupling
of the light into an optical fibre, thereby avoiding the effects of
diffraction. A recent experiment making use of a similar technique is described in \cite{Kurtsiefer2009}.

\begin{figure}[t]
 \centering
    \includegraphics[width=\figwidth]{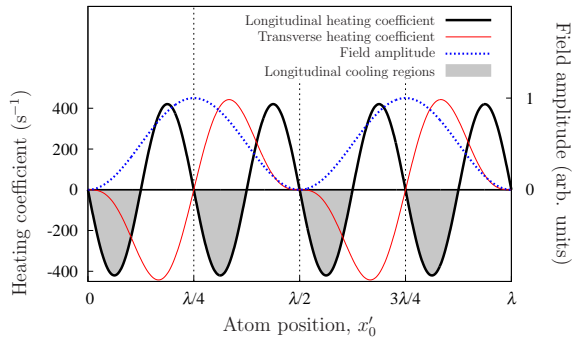}
 \caption{(Color online) Spatial dependence of the longitudinal
 heating coefficient $\heatingcoefft$ (thick solid line). The shaded areas
 promote cooling in the longitudinal direction. Also drawn is the
 transverse heating coefficient (thin solid line) and the field
 amplitude (dotted line). Parameters are for Rb atoms and
 $|A|^2=62.5 \Gamma/(2\pi)$, $\Delta=-10\Gamma$, $\tau=0.25/\Gamma$,
 $w=0.7~\upmu$m.}
 \label{fig:Analytic-Spatial}
\end{figure}

\Eref{eq:AnalyticLongFriction} indicates an exponential decay or
increase in velocity. We define the {\it heating coefficient}
$\heatingcoefft$ of an ensemble of atoms as the proportionality
constant in the relation $\rmd p^2/\rmd t=\heatingcoefft p^2$, which thus
depends on position as $\sin(4k_0x_0)$. Moreover, since $p^2\propto
T$ for a thermal ensemble, we also have $\rmd T/\rmd
t=\heatingcoefft T$. Figure \ref{fig:Analytic-Spatial} shows a plot
of $\heatingcoefft$ against atomic position, where we introduced the
coordinate $x_0^\prime$ relative to the nearest node of the standing
wave pump. It is only in certain intervals that we expect the
longitudinal force to be a damping force, as indicated in this
figure by the shaded regions.

We now derive the friction force in the transverse direction, \ie,
orthogonal to the pump beam. In this case the coupling constant $g$
becomes a function $g(r)$, where $r$ is the coordinate in the
transverse direction. For an atom moving at small constant velocity,
we may write $g[r_0+v(t-t_0)]\approx g(r_0)+v(t-t_0)g^{\prime}(r_0)$
where $g^{\prime}(r)=\rmd g/\rmd r$. Substituting this in
\eref{eq:StartingDE} we can derive an expression for the friction
force, $\force_{\perp}(x_0,t)=-\partial\hat{H}/\partial r$, in the
direction of $r$:
\begin{align*}
 \force_{\perp}(x_0,t_0)=-4\pi\hbar
v\tau&|A|^2\left(\frac{2gg^{\prime}\Delta}{\Delta^2+\Gamma^2}\right)^{\!\!2}
\nonumber\\
&\times\ \sin^3(k_0x_0)\cos(k_0x_0)\text{.}
\end{align*}
This transverse friction force is also shown in
\fref{fig:Analytic-Spatial} assuming a Gaussian mode function of
waist $w=0.7\upmu$m. Note that $\force_{\parallel}$ and
$\force_{\perp}$ are comparable in magnitude for the parameters
chosen here where the mode waist is comparable to the optical
wavelength, and that there exist regions where both these forces
promote cooling.

In the remainder of this paper, however, we will concentrate on a
one-dimensional treatment of the problem and therefore only consider
the longitudinal friction force. This could correspond, for example,
to the imaging arrangement of Eschner at al.\ \cite{Eschner2001}.

In terms of more familiar parameters, we can
rewrite~\eref{eq:AnalyticLongFriction} in the limit
$|\Delta|\gg\Gamma$ as
\begin{equation}
\label{eq:AnalyticLongFrictionFamiliar}
 \force_{\parallel}(x_0,t)=-4vs\Gamma\frac{\sigma_{\text{a}}}{\pi w^2}\hbar
k_0^2\tau\sin(4k_0x_0)\text{,}
\end{equation}
where $s=g^2|A|^2/(\Delta^2+\Gamma^2)$ is the maximum saturation
parameter of the atom in the standing wave,
$\sigma_{\text{a}}=3\lambda^2/(2\pi)$ is the atomic radiative
cross-section, and where we used the relation $2\pi g^2/\Gamma=4
\sigma_a/(\pi w^2)$.

Aside from allowing us to make predictions of cooling
times,~\eref{eq:AnalyticLongFriction}
and~\eref{eq:AnalyticLongFrictionFamiliar} also highlight the
dependence of this cooling effect on the variation of certain
physical parameters. In particular, $\force_{\parallel}$ depends on
the square of the detuning, which means that it is possible to
obtain cooling with both positive and negative detuning. The
friction force also scales with $w^{-4}$ and $|A|^2$. Hence, for a
fixed laser intensity, proportional to $|A|^2/w^2$, \ie, fixed
atomic saturation, friction still scales with $w^{-2}$ and thus a
tight focus is needed in order to have a sizeable effect. A very
promising feature of these two equations is the linear dependence of
the cooling rate on $\tau$. In~\sref{sec:Numerical} we further
analyze the dependence of the cooling rate on the various parameters
and support the validity of the analytic solution by comparing it
with the results of simulations.

\subsection{Localizing the particle}
\label{sec:Perturbative:Dipole}

\Eref{eq:AnalyticLongFriction} shows that, in order to observe any
cooling effects, we need to localize the particle within around
$\lambda/8$. This can be achieved, for example, by an additional
far-off resonant and tightly focused laser beam propagating parallel
to the mirror forming a dipole trap centered at a point $x_0$. We
characterize this trap by means of its spring constant $k_t$, such
that the trapping force is given by $F_t=-k_t(x-x_0)$, or
equivalently by the harmonic oscillator frequency
$\omega_{\text{t}}=\sqrt{(k_t/m)}$, where $m$ is the mass of the atom.

If we now assume that the atom oscillates as
$x(t)=x_0+x_{\text{m}}\sin[\omega_{\text{t}}(t-t_0)]$ in the trap
with a maximum distance $x_{\text{m}}$ of the particle from the trap
center and a corresponding maximum velocity $v_{\text{m}}$, it is
possible to derive a new friction coefficient by perturbation theory
in $x_{\text{m}}$. Proceeding along the lines of
\sref{sec:PerturbativeFriction}, we arrive at
\begin{align}
\label{eq:AnalyticLongFrictionHarm}
 \force_{\parallel}(x_0,t)=\ &-2\pi\hbar
k_0^2v_m\tau\sinc(2\omega_{\text{t}}\tau)\nonumber\\
&\times\
|A|^2\left(\frac{g^2\Delta}{\Delta^2+\Gamma^2}\right)^{\!\!2}\sin(4k_0x_0)\text{
.}
\end{align}
Note that this formula reduces to~\eref{eq:AnalyticLongFriction} in
the limit of small $w_{\text{t}}$. The sinusoidal dependence on
$\omega_{\text{t}}\tau$ can be explained in an intuitive manner: the
effect on the particle is unchanged if the particle undergoes an
integer number of oscillations in the round-trip time $2\tau$.

While \eref{eq:AnalyticLongFrictionHarm} was derived for an
oscillating particle, it is still only correct to lowest order in
$x_m$ and therefore does not include the effect of a finite spatial
distribution. In order to obtain an estimate for the friction force
in the presence of spatial broadening, we calculate the overall
energy loss rate experienced by the particle in terms of the time average
of~\eref{eq:AnalyticLongFriction}:
\begin{align}
\label{eq:AnalyticFrictionTimeAvg}
 \left\langle\frac{\rmd p^2}{\rmd t}\right\rangle=&\ -\frac{2\hbar
k_0^2p_0^2}{m}\tau|A|^2\left(\frac{g^2\Delta}{\Delta^2+\Gamma^2}\right)^{\!\!2}
\nonumber\\
\times\int_0^{2\pi}&\sin[4k_0x_0+4k_0x_{\text{m}}\sin(T)]\cos^2(T)\,\rmd
T\text{,}
\end{align}
where $p_0=m v_m$ is the maximum momentum of the particle in the
trap given by $p_0=x_{\text{m}}\sqrt{m k_{\text{t}}}$. The value of
the integral in (\ref{eq:AnalyticFrictionTimeAvg}) can be expressed
as
\begin{equation*}
\frac{2\pi}{4k_0x_{\text{m}}}[\sin(4k_0x_0)J_1(4k_0x_{\text{m}}
)+\cos(4k_0x_0)H_1(4k_0x_{\text{m}})],
\end{equation*}
where $J_1$ is the order-$1$ Bessel function of the first kind and
$H_1$ is the order-$1$ Struve function \cite{Gradshteyn1994}. At the
point of maximum friction, $x_0^\prime=-3\lambda/16$, the integral in
the above equation reduces to $2\pi
J_1(4k_0x_{\text{m}})/(4k_0x_{\text{m}})$ which can be readily
evaluated.

For small values of $x_{\text{m}}$, the effect of this averaging
process is to introduce a factor of one half into
\eref{eq:AnalyticLongFrictionHarm}, which can be seen as being
equivalent to the effect of cooling merely one degree of freedom
when the atom is in a harmonic trap. Finally,
\eref{eq:AnalyticFrictionTimeAvg} is modified similarly to
\eref{eq:AnalyticLongFrictionHarm} to include the effects of the
harmonic oscillation by replacing
$\tau\rightarrow\sin(2\omega_{\text{t}}\tau)/(2\omega_{\text{t}})$.
This results in an approximate expression for the friction, taking
into account the periodicity in the time delay as well as spatial
averaging effects,
\begin{align}
\label{eq:AnalyticLongFrictionHarmReduced}
 \left\langle \force_{\parallel}(x_0,t) \right\rangle=&\ -{\hbar
k_0^2v_m}\tau|A|^2\left(\frac{g^2\Delta}{\Delta^2+\Gamma^2}\right)^{\!\!2}
\sinc(2\omega_{\text{t}}\tau)\nonumber\\
\times\int_0^{2\pi}&\sin[4k_0x_0+4k_0x_{\text{m}}\sin(T)]\cos^2(T)\,\rmd
T\text{.}
\end{align}

\subsection{Capture range}
\label{sec:Perturbative:Cutoff}

\begin{figure}[t]
 \centering
    \includegraphics[width=\figwidth]{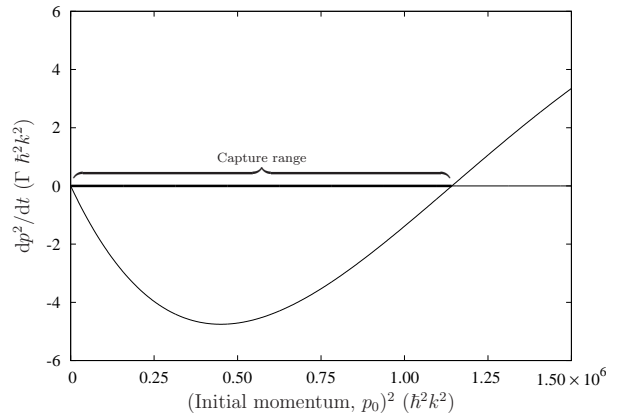}
 \caption{Dependence of the heating rate ($\rmd p^2/\rmd t$) on the
 square of the initial momentum, $p_0^2$, for
 $\omega_{\text{t}}=0.45\times2\pi\Gamma$ and $x_0^\prime=3\lambda/16$.
 Other parameters are as in \fref{fig:Analytic-Spatial}.
 Cooling is achieved only for a finite range of initial momenta.}
 \label{fig:CutoffExpln}
\end{figure}

As discussed above, the addition of the dipole trap introduces
several features into the friction force. Plotting the variation of
the friction force in~\eref{eq:AnalyticFrictionTimeAvg} with the
particle's initial momentum, as in~\fref{fig:CutoffExpln}, shows
that the force changes sign for high enough initial momentum. This
is due to the broader spatial distribution for faster particles in
the harmonic trap. For fast enough velocities, the particle
oscillates into the heating regions, as shown in \fref{fig:Analytic-Spatial},
even if the trap is centered at the
position of maximum cooling. This defines a range of initial
momenta, starting from zero, within which a particle is cooled by
this mechanism; faster particles are heated and ejected from the
trap. Note that this result was derived from the friction to lowest
order in velocity $v$, and higher order terms are expected to affect
the capture range further.

At particular values of $x_0^\prime$, e.g. at $-3\lambda/16$, this
capture range can be conveniently estimated by using the location of
the first zero of the Bessel function,
\begin{equation}
\label{eq:CaptureRangeBesselZero}
p_0 \approx 0.958\sqrt{(m k_t)}/k_0 = 0.958 m \omega_t/k_0.
\end{equation}
Thus, $p_0^2 \propto \omega_{\text{t}}^2$, and the capture range as
defined in \fref{fig:CutoffExpln} is expected to scale with the
square of the trap frequency. We compare this later in \sref{sec:Numerical} with
the results of numerical simulations.

\subsection{Diffusion and steady-state temperature}
\label{sec:SST}

\begin{figure}[t]
 \centering
    \includegraphics[width=\figwidth]{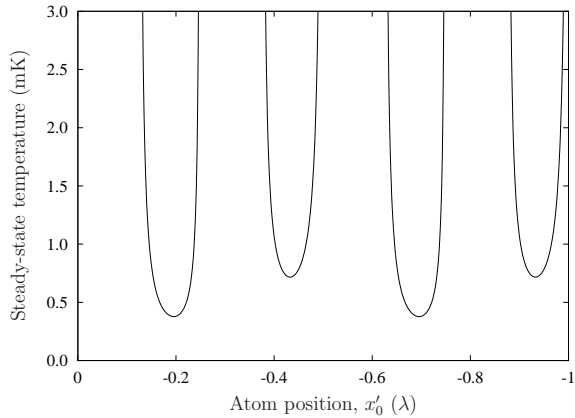}
 \caption{Calculated steady-state temperature $T_{\text{M}}$
 for an atom confined in a harmonic trap as a function of
 position whilst keeping the detuning and pump field constant.
 $\omega_{\text{t}}=0.1\times 2\pi\Gamma$; other parameters are as in
 \fref{fig:Analytic-Spatial}.}
 \label{fig:SST_vs_Posn}
\end{figure}

In the preceding discussion we found a friction force which cools an
atom towards zero momentum. In practice, the cooling process is
counteracted by momentum diffusion due to spontaneous scattering by
the atom of photons from the pump beam into other electromagnetic
modes and between the two counterpropagating components of the
standing wave pump itself. In a simplified Brownian motion model,
this diffusion introduces a constant in the equation for $\rmd
p^2/\rmd t$, resulting in a constant upward shift of the curve
in~\fref{fig:CutoffExpln}. This slightly reduces the capture range
for fast particles, but its main effect is to introduce a specific
value of the momentum where friction and diffusion exactly
compensate each other. This point corresponds to the steady-state
temperature achievable through the cooling mechanism discussed here.

To lowest order in the coupling coefficient $g^2$, the diffusion is
given by the interaction of the atom with the unperturbed,
standing-wave pump field. In this limit, diffusion in our system is
therefore identical to that of Doppler cooling \cite{Gordon1980,
Cook1980, CohenTannoudji1992, BergSorensen1992}, where the diffusion
coefficient $\diffn$ is given to lowest order in $s$ by
\begin{equation}
\label{eq:Diffn}
\diffn=\hbar^2k_0^2\Gamma
s\Big[\cos^2(k_0x_0)+\tfrac{2}{5}\sin^2(k_0x_0)\Big]\text{.}
\end{equation}

The steady-state temperature $T_M$ of mirror-mediated cooling is
then obtained from $k_BT_M = \diffn v_m/ \force_{\parallel}(x_0,t)$
where $\force_{\parallel}(x_0,t)$ is the friction force given by
\eref{eq:AnalyticLongFrictionHarm}. For $|\Delta|>>\Gamma$ we
find
\begin{equation}
\label{eq:TempM}
T_{\text{M}}=\frac{1}{5\pi}\frac{\hbar}{k_B}\frac{\omega_{\text{t}}\Gamma}{g^2}
\frac{2+3\cos^2(k_0x_0)}{\sin(2\omega_{\text{t}}\tau)\sin(4k_0x_0)}.
\end{equation}
An example of the dependence of $T_{\text{M}}$ on the trap position
is shown in~\fref{fig:SST_vs_Posn}, predicting a minimum temperature
of the order of $400~\upmu$K. Whilst this may seem large in comparison
to the Doppler temperature of $141~\upmu$K, one has to keep in mind
that $T_{\text{M}}$, given by \eref{eq:TempM}, is insensitive
to detuning and, for far off-resonant operation of the order of tens
of linewidths, it will be the dominant mechanism. This is further discussed in
\sref{sec:Beyond}. We also note that
\fref{fig:SST_vs_Posn} further highlights the importance of the
requirement for localizing the particle.
Using~\eref{eq:AnalyticLongFrictionFamiliar} we can approximate the
steady-state temperature at the point of maximum friction by
\begin{equation*}
 k_BT_{\text{M}}\approx\frac{\hbar}{\tau}\frac{\pi
w^2}{8\sigma_{\text{a}}}\text{.}
\end{equation*}
It is interesting to note that this expression is closely related to
the expression for the limiting temperature in Doppler cooling,
$k_BT=\hbar\Gamma$, but where $\Gamma$ is replaced by the inverse of
the atom--mirror delay time, $1/\tau$, and where a geometrical
factor related to the mode area divided by the atomic cross section
is included.

\section{Numerical simulations\label{sec:Numerical}}

In this section we now investigate a more accurate numerical
model to corroborate the simplified analytical results obtained
above. In order to render the problem numerically tractable, the
continuum of modes is replaced by a discrete set of modes with
frequencies $\omega_k$, $k=1,\dots,N$. The master equation
(\ref{eq:master}) is then converted by use of the Wigner transform
into a Fokker-Planck equation for the atomic and field variables.
Applying a semiclassical approximation and restricting the equation
of motion to second-order derivatives, one arrives at an equivalent
set of stochastic differential equations for a single atom with
momentum $p$ and position $x$ in a discrete multimode field with
mode amplitudes $\alpha_k$ \cite{Horak2001},
\begin{subequations}
\begin{align}
\label{eq:SDE}
 \rmd x=\ &\frac{p}{m}\rmd t\text{,}\\
 \rmd p=\ &i\gamma\big[\mathcal{E}(x)\tfrac{\rmd}{\rmd
x}\mathcal{E}^{\star}(x)-\mathcal{E}^{\star}(x)\tfrac{\rmd}{\rmd
x}\mathcal{E}(x)\big]\rmd t\nonumber\\
&-U_0\big[\mathcal{E}(x)\tfrac{\rmd}{\rmd
x}\mathcal{E}^{\star}(x)+\mathcal{E}^{\star}(x)\tfrac{\rmd}{\rmd
x}\mathcal{E}(x)\big]\rmd t\nonumber\\
&-k_{\text{t}}(x-x_{\text{t}})\rmd t+\rmd P\text{,}\\
 \rmd\alpha_k=\ &i\Delta_k\alpha_k\rmd
t-(iU_0+\gamma)\mathcal{E}(x)f_k^{\star}(x)\rmd
t+\rmd A_k\text{,}
\end{align}
\end{subequations}
where $f_k(x)=\sin(\omega_k x/c)$ are the mode functions,
$\mathcal{E}(x)=\sum_k \alpha_k f_k(x)$ is the total electric field,
$\Delta_k=\omega_0-\omega_k$ is the detuning of each mode from the
pump, $U_0$ is the light shift per photon, and $\gamma$ is the
photon scattering rate. The terms $\rmd P$ and $\rmd A_k$ are
correlated noise terms \cite{Horak2001} responsible for momentum and
field diffusion.

In the following, we set the trap center to
$x_0^\prime=-3\lambda/16$, which is the point where the analytic
solution predicts the maximum of the damping force. We use $N=256$
field modes with a mode spacing of $\Gamma/10$. At the start of
every simulation, all field modes are empty with the exception of
the pump mode which is initialized at $625$ photons, corresponding
to a laser power of around $50$~pW for our chosen parameters.

The simulations were performed in runs of several thousand
trajectories. Each such run was performed at a well-defined initial
temperature, with the starting momenta of the particles chosen from
a Gaussian distribution, and the starting position being the center
of the trap.

\subsection{Friction force and capture range\label{sec:Numerical:Friction}}

\begin{figure}[t]
 \centering
    \subfigure[]{
    \includegraphics[width=0.8\figwidth]{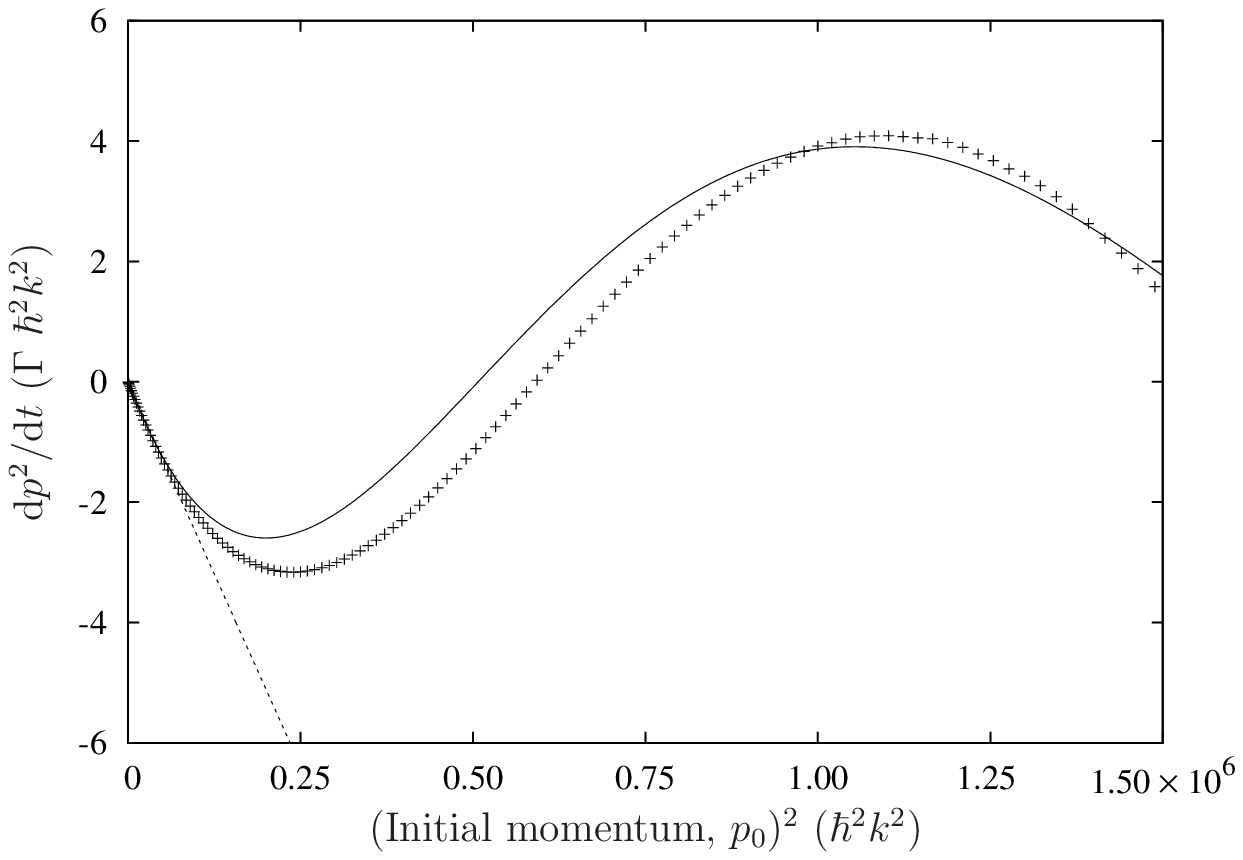}
    }
    \subfigure[]{
    \includegraphics[width=0.82\figwidth]{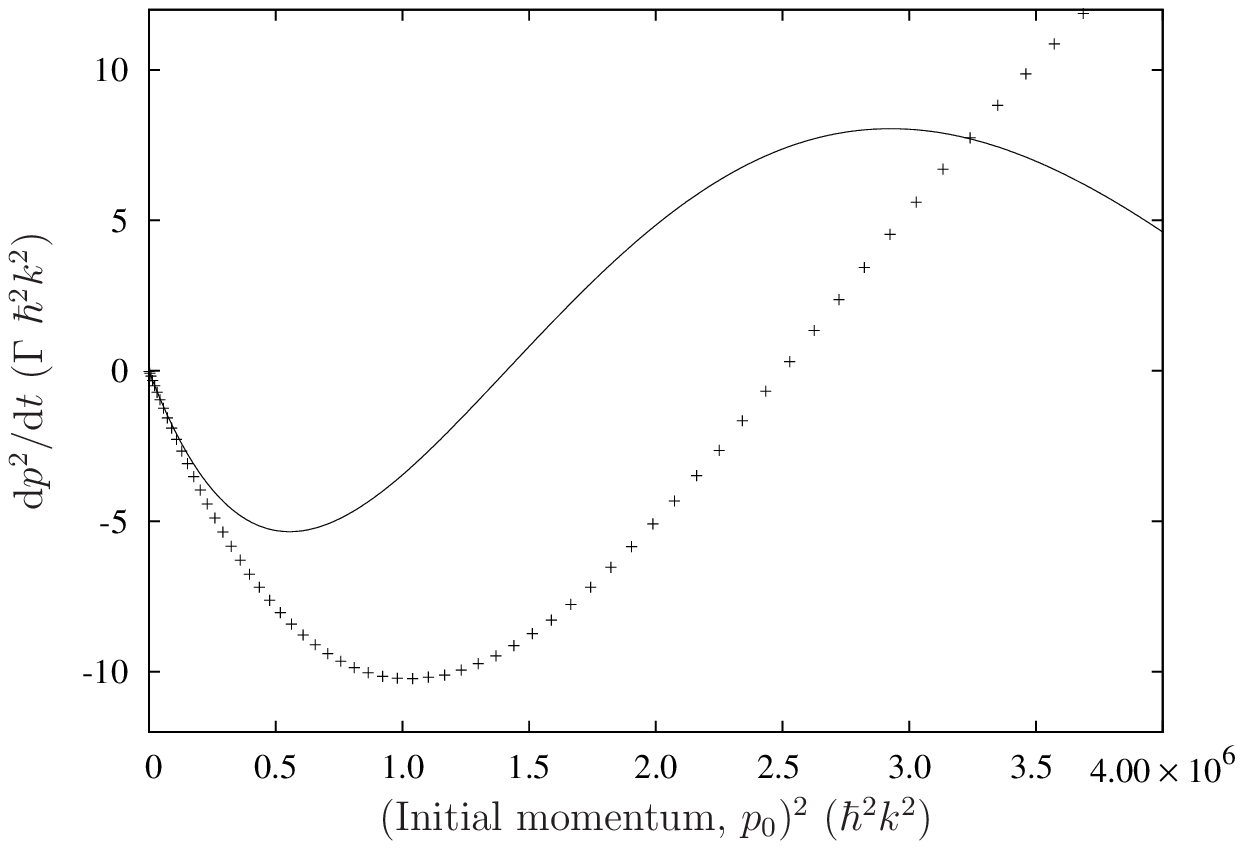}
    }
    \hspace{0.1in}
 \caption{Comparison of heating rate ($\rmd p^2/\rmd t$) for the
 simulations without noise (data points) with the analytic
 approximation, \eref{eq:AnalyticLongFrictionHarmReduced},
 including the harmonic trap (solid line). (a) Weak harmonic
 trap, $\omega_{\text{t}}=0.3\times 2\pi\Gamma$, showing also the
 linear dependence in the limit of small momenta,
\eref{eq:AnalyticLongFrictionHarm} (dotted line). (b) Stiff trap,
 $\omega_{\text{t}}=0.5\times 2\pi\Gamma$. The trap position
 $x_0^\prime=-3\lambda/16$ and other parameters are as in
\fref{fig:Analytic-Spatial}.}
 \label{fig:Comparison}
\end{figure}

\Fref{fig:Comparison} presents the results of a set of simulations
performed when setting the noise terms $\rmd P$ and $\rmd A_k$ in
equations (\ref{eq:SDE}) to zero, \ie, neglecting momentum and
photon number diffusion. The simulation data are compared with the
result of the perturbative calculations
\eref{eq:AnalyticLongFrictionHarmReduced}. For modest values of
$\omega_{\text{t}}$, \fref{fig:Comparison}(a) justifies the
averaging process used to derive (\ref{eq:AnalyticFrictionTimeAvg})
which was based on spatial averaging but neglecting higher order
terms in $v$. In contrast, for larger trap frequencies, the
numerical simulations diverge significantly from the analytic
result, as can be seen in \fref{fig:Comparison}(b). We expect
that the terms in higher powers of the initial speed, which were
dropped in the perturbative solution, are responsible for this
discrepancy.

\begin{figure}[t]
 \centering
    \includegraphics[width=\figwidth]{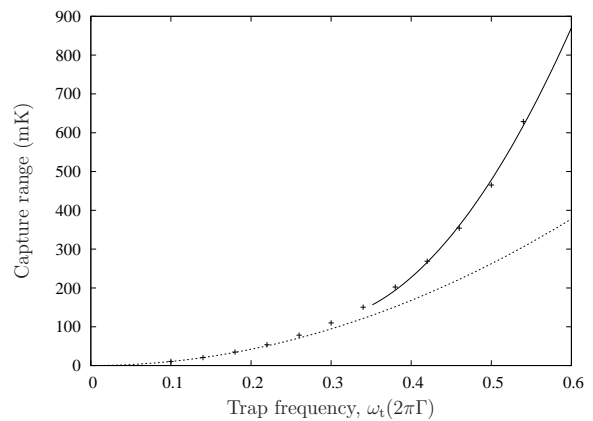}
 \caption{Capture range extracted from the simulations (data points)
 as compared to the analytic solution (dotted line) for
 various values of $\omega_{\text{t}}$. The solid line is a
 quadratic fit to the data for $\omega_{\text{t}}\geq 0.3\times
 2\pi\Gamma$ and is only intended as a guide to the eye.
 Other parameters are as in \fref{fig:Comparison}.}
 \label{fig:Cutoff}
\end{figure}

We have already seen, in \eref{eq:CaptureRangeBesselZero}, that the
capture range is expected to scale as $\omega_{\text{t}}^2$. For
weak traps, as shown in~\fref{fig:Cutoff}, the numerical simulations
agree well with these expectations. For stiffer traps, however, the
capture range is consistently larger than that predicted; in fact,
the simulations predict a capture range of around $450$~mK for a
trap frequency of $0.5\times 2\pi\Gamma$

\subsection{Steady-state temperature\label{sec:Numerical:SST}}

\begin{figure}[t]
 \centering
    \includegraphics[width=\figwidth]{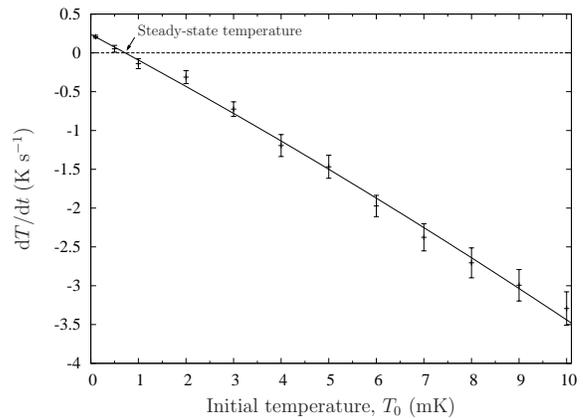}
 \caption{Heating rate ($\rmd T/\rmd t$) extracted from the
 simulations starting at a number of initial temperatures. The solid
 line  represents a quadratic fit to the data. $\omega_t=0.5 \times
 2\pi\Gamma$; other parameters are as in \fref{fig:Comparison}.}
 \label{fig:Heating-vs-Temp}
\end{figure}

The next step in our investigation was to run simulations involving
the full dynamics given by Eqs.\ (\ref{eq:SDE}) including the
diffusion terms. Because of the discrete nature of the field modes
with uniform frequency spacing used in the simulations, the
numerically modeled behavior is always periodic in time with a
periodicity given by the inverse of the frequency spacing. The
simulations therefore cannot follow each trajectory to its
steady-state. Instead, simulations were performed in several groups
of trajectories, each group forming a thermal ensemble at a
well-defined initial temperature. For each such group of
trajectories the initial value of $\rmd T/\rmd t$ was calculated.
The results for $\omega_{\text{t}}=0.5\times 2\pi\Gamma$ are shown
in \fref{fig:Heating-vs-Temp}, where the error bars are due to
statistical fluctuations for a finite number of stochastic
integrations. The steady-state temperature is that temperature at
which $\rmd T/\rmd t=0$ as clearly illustrated in this figure. For
the chosen parameters, our data suggest a steady-state temperature
of $722\pm 54$~$\upmu$K with a $1/e$ cooling time of around $3.0$~ms.
This compares reasonably well with the steady-state temperature of
$597$~$\upmu$K predicted by \eref{eq:TempM}.

\begin{figure}[t]
 \centering
    \includegraphics[width=\figwidth]{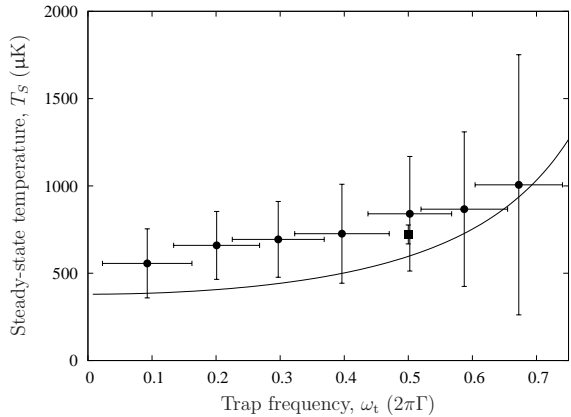}
 \caption{Steady-state temperature for a number of simulations
 (circles) compared to the analytic formula (\ref{eq:TempM}) (solid
 line). The solid square represents the equivalent data
 from~\fref{fig:Heating-vs-Temp}, resulting from a much larger
 number of simulations. Parameters are as in \fref{fig:Comparison}. }
 \label{fig:SST}
\end{figure}

We finally performed a large number of simulations to investigate
the dependence of the steady-state temperature on the trap
frequency. \Eref{eq:TempM} indicates that as one decreases
$\omega_{\text{t}}$ the steady state temperature decreases. This is
clearly seen in~\fref{fig:SST}, which compares the prediction
of~\eref{eq:TempM} with a set of numerical simulations. The trend in
the data is reproduced well by the analytic expression. However, the
simulated steady-state temperature is consistently a little higher than
predicted. We expect that this discrepancy is due to one of two
reasons. (i) Equation (\ref{eq:TempM}) was derived from the friction~\eref{eq:AnalyticLongFrictionHarm}, \ie, without the spatial
averaging of (\ref{eq:AnalyticLongFrictionHarmReduced}) which would
reduce friction. (ii) Higher order terms in the velocity $v$ are
also expected to reduce friction compared to the lowest order
analytical result. In both cases, therefore, the analytic expression
is expected to overestimate the friction force and thus
to predict too low temperatures.

\section{Beyond adiabatic theory\label{sec:Beyond}}

\begin{figure}[t]
 \centering
    \includegraphics[width=\figwidth]{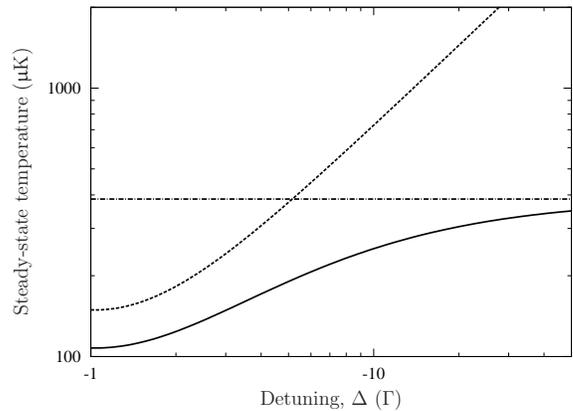}
 \caption{Comparison between the calculated steady-state
 temperatures for mirror-mediated cooling $T_{\text{M}}$ (dash-dotted line),
 Doppler cooling $T_{\text{D}}$ (dashed), and in the presence of
 both effects $T$ (solid), drawn as a function of detuning whilst keeping the
 saturation parameter constant. $\omega_{\text{t}}=0.1\times
 2\pi\Gamma$; other parameters are as in \fref{fig:Comparison}.}
 \label{fig:Doppler_vs_Mirror}
\end{figure}

All the theoretical analysis and simulations discussed so far have
been based on adiabatic elimination of the internal atomic degrees
of freedom, and therefore neglected Doppler cooling.
In~\fref{fig:Doppler_vs_Mirror}, we explore the variation of
$T_{\text{M}}$ and the Doppler temperature, $T_{\text{D}}$, as a
function of detuning from resonance when the particle is at the
point of greatest friction ($x_0^\prime=3\lambda/16$), where $T_D$
is given by $T_D = -\hbar\Gamma (\Delta^2+\Gamma^2)/(2\Delta)$ for
negative values of $\Delta$. In the presence of both cooling
effects, the stationary temperature achieved by the system is given
by
 \begin{equation}
 T = \left( \frac{1}{T_{\text{M}}}+\frac{1}{T_{\text{D}}}\right)^{-1}.
 \label{eq:fulltemp}
 \end{equation}
Thus, for the parameters of \fref{fig:SST}, the calculated
steady-state temperature $T$ reduces to $250$~$\upmu$K in the limit of
vanishing $\omega_{\text{t}}$.

From~\fref{fig:Doppler_vs_Mirror} one can see that the
mirror-mediated force, for our tightly focused pump, is stronger
than the Doppler force for detunings larger than around $10\Gamma$
in magnitude. In practice this has two implications: for large
negative detunings, we expect the steady-state temperature of the
system to be significantly lower than that predicted by Doppler
cooling; whereas for large \emph{positive} detunings, we still
predict equilibrium temperatures of the order of mK.

Both our perturbative expressions and our simulations are calculated
to lowest orders in the atomic saturation. However, it is well known
that in the limit of very large detunings also higher order terms in
the saturation parameter $s$ become significant. Using the full
expression for the diffusion constant \cite{Gordon1980}, we can
estimate the detuning for which we expect minimum diffusion and
temperature. For the value of the saturation parameter $s\lesssim
0.1$ used throughout this paper, it can be shown that $T_{\text{M}}$
attains a minimum at detunings of up to several tens of linewidths.
Our chosen parameters are therefore within the range of validity of
the model.

\section{Conclusion\label{sec:Conclusion}}

We have presented a mechanism for cooling particles by optical
means which is based fundamentally on the dipole interaction of a
particle with a light beam and therefore does not rely on
spontaneous emission. The particle is assumed to be trapped and is
simultaneously driven by an off-resonant laser beam. After the
interaction with the particle the beam is reflected back onto the
particle by a distant mirror. The time-delay incurred during the
light round-trip to the mirror and back is exploited to create a
non-conservative cooling force.

The system was analyzed using stochastic simulations of the
semiclassical equations of motion representing a single two-level
atom coupled to a continuum of electromagnetic modes. The results of
these computations were found to agree with the expectations of a
perturbative analysis. Our models predict sub-mK steady-state
temperatures for $^{85}$Rb atoms interacting with a tightly focused
laser beam several meters from the mirror, in an arrangement similar
to that of Ref.\ \cite{Eschner2001}. While most of the theory is
presented for a one-dimensional model, results for the friction
force in the transverse direction suggest that three-dimensional
cooling is possible with this scheme.

The model presented here requires a large separation between the
atom and the mirror, of the order of several meters, for an
observable cooling effect. This limitation can be overcome in
several ways. First, the light could propagate in an optical fiber
between the atom and the mirror to avoid the effects of diffraction.
Second, the required delayed reflection could be achieved through
the use of a cavity instead of a mirror; in contrast to
cavity-mediated cooling schemes \cite{Horak1997, Vuletic2000,
Maunz2004, Vilensky2007, Lev2008}, the atom would remain external to
the cavity. For a time delay $\tau$ of order $1$~ns one would
require a cavity quality factor $Q=\omega\tau$ \cite{Rempe1992} of
the order of $10^6-10^7$, which is achievable with present-day
technology \cite{Mabuchi1994}.

\begin{acknowledgments}
The authors thank Peter Domokos and Helmut Ritsch for helpful
discussions. This work was supported by the UK Engineering and
Physical Sciences Research Council (EPSRC) grant EP/E058949/1 and by
the Cavity-Mediated Molecular Cooling network within the EuroQUAM
programme of the European Science Foundation (ESF).
\end{acknowledgments}


\begin{thebibliography}{36}
\expandafter\ifx\csname natexlab\endcsname\relax\def\natexlab#1{#1}\fi
\expandafter\ifx\csname bibnamefont\endcsname\relax
  \def\bibnamefont#1{#1}\fi
\expandafter\ifx\csname bibfnamefont\endcsname\relax
  \def\bibfnamefont#1{#1}\fi
\expandafter\ifx\csname citenamefont\endcsname\relax
  \def\citenamefont#1{#1}\fi
\expandafter\ifx\csname url\endcsname\relax
  \def\url#1{\texttt{#1}}\fi
\expandafter\ifx\csname urlprefix\endcsname\relax\def\urlprefix{URL }\fi
\providecommand{\bibinfo}[2]{#2}
\providecommand{\eprint}[2][]{\url{#2}}

\bibitem[{\citenamefont{H\"{a}nsch and Schawlow}(1975)}]{Hansch1975}
\bibinfo{author}{\bibfnamefont{T.~W.} \bibnamefont{H\"{a}nsch}}
  \bibnamefont{and} \bibinfo{author}{\bibfnamefont{A.~L.}
  \bibnamefont{Schawlow}}, \bibinfo{journal}{Opt. Commun.}
  \textbf{\bibinfo{volume}{13}}, \bibinfo{pages}{68} (\bibinfo{year}{1975}).

\bibitem[{\citenamefont{Wineland and Itano}(1979)}]{Wineland1979}
\bibinfo{author}{\bibfnamefont{D.~J.} \bibnamefont{Wineland}} \bibnamefont{and}
  \bibinfo{author}{\bibfnamefont{W.~M.} \bibnamefont{Itano}},
  \bibinfo{journal}{Phys. Rev. A} \textbf{\bibinfo{volume}{20}},
  \bibinfo{pages}{1521} (\bibinfo{year}{1979}).

\bibitem[{\citenamefont{Mellish and Wilson}(2002)}]{Mellish2002}
\bibinfo{author}{\bibfnamefont{A.~S.} \bibnamefont{Mellish}} \bibnamefont{and}
  \bibinfo{author}{\bibfnamefont{A.~C.} \bibnamefont{Wilson}},
  \bibinfo{journal}{Am. J. Phys.} \textbf{\bibinfo{volume}{70}},
  \bibinfo{pages}{965} (\bibinfo{year}{2002}).

\bibitem[{\citenamefont{Gabbanini et~al.}(2000)\citenamefont{Gabbanini,
  Fioretti, Lucchesini, Gozzini, and Mazzoni}}]{Gabbanini2000}
\bibinfo{author}{\bibfnamefont{C.}~\bibnamefont{Gabbanini}},
  \bibinfo{author}{\bibfnamefont{A.}~\bibnamefont{Fioretti}},
  \bibinfo{author}{\bibfnamefont{A.}~\bibnamefont{Lucchesini}},
  \bibinfo{author}{\bibfnamefont{S.}~\bibnamefont{Gozzini}}, \bibnamefont{and}
  \bibinfo{author}{\bibfnamefont{M.}~\bibnamefont{Mazzoni}},
  \bibinfo{journal}{Phys. Rev. Lett.} \textbf{\bibinfo{volume}{84}},
  \bibinfo{pages}{2814} (\bibinfo{year}{2000}).

\bibitem[{\citenamefont{Sage et~al.}(2005)\citenamefont{Sage, Sainis, Bergeman,
  and DeMille}}]{Sage2005}
\bibinfo{author}{\bibfnamefont{J.~M.} \bibnamefont{Sage}},
  \bibinfo{author}{\bibfnamefont{S.}~\bibnamefont{Sainis}},
  \bibinfo{author}{\bibfnamefont{T.}~\bibnamefont{Bergeman}}, \bibnamefont{and}
  \bibinfo{author}{\bibfnamefont{D.}~\bibnamefont{DeMille}},
  \bibinfo{journal}{Phys. Rev. Lett.} \textbf{\bibinfo{volume}{94}},
  \bibinfo{pages}{203001} (\bibinfo{year}{2005}).

\bibitem[{\citenamefont{Winkler et~al.}(2007)\citenamefont{Winkler, Lang,
  Thalhammer, v.~d. Straten, Grimm, and Denschlag}}]{Winkler2007}
\bibinfo{author}{\bibfnamefont{K.}~\bibnamefont{Winkler}},
  \bibinfo{author}{\bibfnamefont{F.}~\bibnamefont{Lang}},
  \bibinfo{author}{\bibfnamefont{G.}~\bibnamefont{Thalhammer}},
  \bibinfo{author}{\bibfnamefont{P.}~\bibnamefont{v.~d. Straten}},
  \bibinfo{author}{\bibfnamefont{R.}~\bibnamefont{Grimm}}, \bibnamefont{and}
  \bibinfo{author}{\bibfnamefont{J.~H.} \bibnamefont{Denschlag}},
  \bibinfo{journal}{Phys. Rev. Lett.} \textbf{\bibinfo{volume}{98}},
  \bibinfo{pages}{043201} (\bibinfo{year}{2007}).

\bibitem[{\citenamefont{Metcalf and van~der Straten}(2003)}]{Metcalf2003}
\bibinfo{author}{\bibfnamefont{H.~J.} \bibnamefont{Metcalf}} \bibnamefont{and}
  \bibinfo{author}{\bibfnamefont{P.}~\bibnamefont{van~der Straten}},
  \bibinfo{journal}{J. Opt. Soc. Am. B} \textbf{\bibinfo{volume}{20}},
  \bibinfo{pages}{887} (\bibinfo{year}{2003}).

\bibitem[{\citenamefont{Pinkse et~al.}(2003)\citenamefont{Pinkse, Junglen,
  Rieger, Rangwala, and Rempe}}]{Pinkse2003}
\bibinfo{author}{\bibfnamefont{P.~W.~H.} \bibnamefont{Pinkse}},
  \bibinfo{author}{\bibfnamefont{P.~T.} \bibnamefont{Junglen}},
  \bibinfo{author}{\bibfnamefont{T.}~\bibnamefont{Rieger}},
  \bibinfo{author}{\bibfnamefont{S.~A.} \bibnamefont{Rangwala}},
  \bibnamefont{and} \bibinfo{author}{\bibfnamefont{G.}~\bibnamefont{Rempe}}, in
  \emph{\bibinfo{booktitle}{European Quantum Electronics Conference (EQEC
  2003), Munich, Germany}} (\bibinfo{publisher}{IEEE}, \bibinfo{year}{2003}),
  p. \bibinfo{pages}{271}.

\bibitem[{\citenamefont{Kerman et~al.}(2000)\citenamefont{Kerman, Vuletic,
  Chin, and Chu}}]{Kerman2000}
\bibinfo{author}{\bibfnamefont{A.~J.} \bibnamefont{Kerman}},
  \bibinfo{author}{\bibfnamefont{V.}~\bibnamefont{Vuletic}},
  \bibinfo{author}{\bibfnamefont{C.}~\bibnamefont{Chin}}, \bibnamefont{and}
  \bibinfo{author}{\bibfnamefont{S.}~\bibnamefont{Chu}},
  \bibinfo{journal}{Phys. Rev. Lett.} \textbf{\bibinfo{volume}{84}},
  \bibinfo{pages}{439} (\bibinfo{year}{2000}).

\bibitem[{\citenamefont{Horak et~al.}(1997)\citenamefont{Horak, Hechenblaikner,
  Gheri, Stecher, and Ritsch}}]{Horak1997}
\bibinfo{author}{\bibfnamefont{P.}~\bibnamefont{Horak}},
  \bibinfo{author}{\bibfnamefont{G.}~\bibnamefont{Hechenblaikner}},
  \bibinfo{author}{\bibfnamefont{K.~M.} \bibnamefont{Gheri}},
  \bibinfo{author}{\bibfnamefont{H.}~\bibnamefont{Stecher}}, \bibnamefont{and}
  \bibinfo{author}{\bibfnamefont{H.}~\bibnamefont{Ritsch}},
  \bibinfo{journal}{Phys. Rev. Lett.} \textbf{\bibinfo{volume}{79}},
  \bibinfo{pages}{4974} (\bibinfo{year}{1997}).

\bibitem[{\citenamefont{Vuletic and Chu}(2000)}]{Vuletic2000}
\bibinfo{author}{\bibfnamefont{V.}~\bibnamefont{Vuletic}} \bibnamefont{and}
  \bibinfo{author}{\bibfnamefont{S.}~\bibnamefont{Chu}},
  \bibinfo{journal}{Phys. Rev. Lett.} \textbf{\bibinfo{volume}{84}},
  \bibinfo{pages}{3787} (\bibinfo{year}{2000}).

\bibitem[{\citenamefont{Maunz et~al.}(2004)\citenamefont{Maunz, Puppe,
  Schuster, Syassen, Pinkse, and Rempe}}]{Maunz2004}
\bibinfo{author}{\bibfnamefont{P.}~\bibnamefont{Maunz}},
  \bibinfo{author}{\bibfnamefont{T.}~\bibnamefont{Puppe}},
  \bibinfo{author}{\bibfnamefont{I.}~\bibnamefont{Schuster}},
  \bibinfo{author}{\bibfnamefont{N.}~\bibnamefont{Syassen}},
  \bibinfo{author}{\bibfnamefont{P.~W.~H.} \bibnamefont{Pinkse}},
  \bibnamefont{and} \bibinfo{author}{\bibfnamefont{G.}~\bibnamefont{Rempe}},
  \bibinfo{journal}{Nature} \textbf{\bibinfo{volume}{428}}, \bibinfo{pages}{50}
  (\bibinfo{year}{2004}).

\bibitem[{\citenamefont{Vilensky et~al.}(2007)\citenamefont{Vilensky, Prior,
  and Averbukh}}]{Vilensky2007}
\bibinfo{author}{\bibfnamefont{M.~Y.} \bibnamefont{Vilensky}},
  \bibinfo{author}{\bibfnamefont{Y.}~\bibnamefont{Prior}}, \bibnamefont{and}
  \bibinfo{author}{\bibfnamefont{I.~S.} \bibnamefont{Averbukh}},
  \bibinfo{journal}{Phys. Rev. Lett.} \textbf{\bibinfo{volume}{99}},
  \bibinfo{pages}{103002} (\bibinfo{year}{2007}).

\bibitem[{\citenamefont{Lev et~al.}(2008)\citenamefont{Lev, Vukics, Hudson,
  Sawyer, Domokos, Ritsch, and Ye}}]{Lev2008}
\bibinfo{author}{\bibfnamefont{B.~L.} \bibnamefont{Lev}},
  \bibinfo{author}{\bibfnamefont{A.}~\bibnamefont{Vukics}},
  \bibinfo{author}{\bibfnamefont{E.~R.} \bibnamefont{Hudson}},
  \bibinfo{author}{\bibfnamefont{B.~C.} \bibnamefont{Sawyer}},
  \bibinfo{author}{\bibfnamefont{P.}~\bibnamefont{Domokos}},
  \bibinfo{author}{\bibfnamefont{H.}~\bibnamefont{Ritsch}}, \bibnamefont{and}
  \bibinfo{author}{\bibfnamefont{J.}~\bibnamefont{Ye}}, \bibinfo{journal}{Phys.
  Rev. A} \textbf{\bibinfo{volume}{77}}, \bibinfo{pages}{023402}
  (\bibinfo{year}{2008}).

\bibitem[{\citenamefont{Folman et~al.}(2002)\citenamefont{Folman, Krueger,
  Schmiedmayer, Denschlag, and Henkel}}]{Folman2002}
\bibinfo{author}{\bibfnamefont{R.}~\bibnamefont{Folman}},
  \bibinfo{author}{\bibfnamefont{P.}~\bibnamefont{Krueger}},
  \bibinfo{author}{\bibfnamefont{J.}~\bibnamefont{Schmiedmayer}},
  \bibinfo{author}{\bibfnamefont{J.}~\bibnamefont{Denschlag}},
  \bibnamefont{and} \bibinfo{author}{\bibfnamefont{C.}~\bibnamefont{Henkel}},
  \bibinfo{journal}{Adv. At. Mol. Opt. Phy.} \textbf{\bibinfo{volume}{48}},
  \bibinfo{pages}{263} (\bibinfo{year}{2002}).

\bibitem[{\citenamefont{Braginsky}(2002)}]{Braginsky2002}
\bibinfo{author}{\bibfnamefont{V.}~\bibnamefont{Braginsky}},
  \bibinfo{journal}{Phys. Lett. A} \textbf{\bibinfo{volume}{293}},
  \bibinfo{pages}{228} (\bibinfo{year}{2002}).

\bibitem[{\citenamefont{Bhattacharya et~al.}(2008)\citenamefont{Bhattacharya,
  Uys, and Meystre}}]{Bhattacharya2008}
\bibinfo{author}{\bibfnamefont{M.}~\bibnamefont{Bhattacharya}},
  \bibinfo{author}{\bibfnamefont{H.}~\bibnamefont{Uys}}, \bibnamefont{and}
  \bibinfo{author}{\bibfnamefont{P.}~\bibnamefont{Meystre}},
  \bibinfo{journal}{Phys. Rev. A} \textbf{\bibinfo{volume}{77}},
  \bibinfo{eid}{033819} (\bibinfo{year}{2008}).

\bibitem[{\citenamefont{Corbitt et~al.}(2007)\citenamefont{Corbitt, Wipf,
  Bodiya, Ottaway, Sigg, Smith, Whitcomb, and Mavalvala}}]{Corbitt2007}
\bibinfo{author}{\bibfnamefont{T.}~\bibnamefont{Corbitt}},
  \bibinfo{author}{\bibfnamefont{C.}~\bibnamefont{Wipf}},
  \bibinfo{author}{\bibfnamefont{T.}~\bibnamefont{Bodiya}},
  \bibinfo{author}{\bibfnamefont{D.}~\bibnamefont{Ottaway}},
  \bibinfo{author}{\bibfnamefont{D.}~\bibnamefont{Sigg}},
  \bibinfo{author}{\bibfnamefont{N.}~\bibnamefont{Smith}},
  \bibinfo{author}{\bibfnamefont{S.}~\bibnamefont{Whitcomb}}, \bibnamefont{and}
  \bibinfo{author}{\bibfnamefont{N.}~\bibnamefont{Mavalvala}},
  \bibinfo{journal}{Phys. Rev. Lett.} \textbf{\bibinfo{volume}{99}},
  \bibinfo{pages}{160801} (\bibinfo{year}{2007}).

\bibitem[{\citenamefont{Schliesser et~al.}(2008)\citenamefont{Schliesser,
  Riviere, Anetsberger, Arcizet, and Kippenberg}}]{Schliesser2008}
\bibinfo{author}{\bibfnamefont{A.}~\bibnamefont{Schliesser}},
  \bibinfo{author}{\bibfnamefont{R.}~\bibnamefont{Riviere}},
  \bibinfo{author}{\bibfnamefont{G.}~\bibnamefont{Anetsberger}},
  \bibinfo{author}{\bibfnamefont{O.}~\bibnamefont{Arcizet}}, \bibnamefont{and}
  \bibinfo{author}{\bibfnamefont{T.~J.} \bibnamefont{Kippenberg}},
  \bibinfo{journal}{Nat Phys} \textbf{\bibinfo{volume}{4}},
  \bibinfo{pages}{415} (\bibinfo{year}{2008}).

\bibitem[{\citenamefont{Xuereb et~al.}(2009)\citenamefont{Xuereb, Domokos,
  Asb\'{\o}th, Horak, and Freegarde}}]{Xuereb2009b}
\bibinfo{author}{\bibfnamefont{A.}~\bibnamefont{Xuereb}},
  \bibinfo{author}{\bibfnamefont{P.}~\bibnamefont{Domokos}},
  \bibinfo{author}{\bibfnamefont{J.}~\bibnamefont{Asb\'{\o}th}},
  \bibinfo{author}{\bibfnamefont{P.}~\bibnamefont{Horak}}, \bibnamefont{and}
  \bibinfo{author}{\bibfnamefont{T.}~\bibnamefont{Freegarde}},
  \bibinfo{journal}{Phys. Rev. A} \textbf{\bibinfo{volume}{79}},
  \bibinfo{pages}{053810} (\bibinfo{year}{2009}).

\bibitem[{\citenamefont{Burns et~al.}(1990)\citenamefont{Burns, Fournier, and
  Golovchenko}}]{Burns1990}
\bibinfo{author}{\bibfnamefont{M.~M.} \bibnamefont{Burns}},
  \bibinfo{author}{\bibfnamefont{J.~M.} \bibnamefont{Fournier}},
  \bibnamefont{and} \bibinfo{author}{\bibfnamefont{J.~A.}
  \bibnamefont{Golovchenko}}, \bibinfo{journal}{Science}
  \textbf{\bibinfo{volume}{249}}, \bibinfo{pages}{749} (\bibinfo{year}{1990}).

\bibitem[{\citenamefont{Mac{d}onald et~al.}(2002)\citenamefont{Mac{d}onald,
  Paterson, Sepulveda, Arlt, Sibbett, and Dholakia}}]{MacDonald2002}
\bibinfo{author}{\bibfnamefont{M.~P.} \bibnamefont{Mac{d}onald}},
  \bibinfo{author}{\bibfnamefont{L.}~\bibnamefont{Paterson}},
  \bibinfo{author}{\bibfnamefont{V.~K.} \bibnamefont{Sepulveda}},
  \bibinfo{author}{\bibfnamefont{J.~J.} \bibnamefont{Arlt}},
  \bibinfo{author}{\bibfnamefont{W.}~\bibnamefont{Sibbett}}, \bibnamefont{and}
  \bibinfo{author}{\bibfnamefont{K.}~\bibnamefont{Dholakia}},
  \bibinfo{journal}{Science} \textbf{\bibinfo{volume}{296}},
  \bibinfo{pages}{1101} (\bibinfo{year}{2002}).

\bibitem[{\citenamefont{Metzger et~al.}(2006)\citenamefont{Metzger, Wright,
  Sibbett, and Dholakia}}]{Metzger2006a}
\bibinfo{author}{\bibfnamefont{N.~K.} \bibnamefont{Metzger}},
  \bibinfo{author}{\bibfnamefont{E.~M.} \bibnamefont{Wright}},
  \bibinfo{author}{\bibfnamefont{W.}~\bibnamefont{Sibbett}}, \bibnamefont{and}
  \bibinfo{author}{\bibfnamefont{K.}~\bibnamefont{Dholakia}},
  \bibinfo{journal}{Opt. Express} \textbf{\bibinfo{volume}{14}},
  \bibinfo{pages}{3677} (\bibinfo{year}{2006}).

\bibitem[{\citenamefont{Eschner et~al.}(2001)\citenamefont{Eschner, Raab,
  Schmidt-Kaler, and Blatt}}]{Eschner2001}
\bibinfo{author}{\bibfnamefont{J.}~\bibnamefont{Eschner}},
  \bibinfo{author}{\bibfnamefont{C.}~\bibnamefont{Raab}},
  \bibinfo{author}{\bibfnamefont{F.}~\bibnamefont{Schmidt-Kaler}},
  \bibnamefont{and} \bibinfo{author}{\bibfnamefont{R.}~\bibnamefont{Blatt}},
  \bibinfo{journal}{Nature} \textbf{\bibinfo{volume}{413}},
  \bibinfo{pages}{495} (\bibinfo{year}{2001}).

\bibitem[{\citenamefont{Bushev et~al.}(2004)\citenamefont{Bushev, Wilson,
  Eschner, Raab, Schmidt-Kaler, Becher, and Blatt}}]{Bushev2004}
\bibinfo{author}{\bibfnamefont{P.}~\bibnamefont{Bushev}},
  \bibinfo{author}{\bibfnamefont{A.}~\bibnamefont{Wilson}},
  \bibinfo{author}{\bibfnamefont{J.}~\bibnamefont{Eschner}},
  \bibinfo{author}{\bibfnamefont{C.}~\bibnamefont{Raab}},
  \bibinfo{author}{\bibfnamefont{F.}~\bibnamefont{Schmidt-Kaler}},
  \bibinfo{author}{\bibfnamefont{C.}~\bibnamefont{Becher}}, \bibnamefont{and}
  \bibinfo{author}{\bibfnamefont{R.}~\bibnamefont{Blatt}},
  \bibinfo{journal}{Phys. Rev. Lett.} \textbf{\bibinfo{volume}{92}},
  \bibinfo{pages}{223602} (\bibinfo{year}{2004}).

\bibitem[{\citenamefont{Gangl and Ritsch}(2000)}]{Gangl2000a}
\bibinfo{author}{\bibfnamefont{M.}~\bibnamefont{Gangl}} \bibnamefont{and}
  \bibinfo{author}{\bibfnamefont{H.}~\bibnamefont{Ritsch}},
  \bibinfo{journal}{Phys. Rev. A} \textbf{\bibinfo{volume}{61}},
  \bibinfo{pages}{043405} (\bibinfo{year}{2000}).

\bibitem[{\citenamefont{Gardiner}(1984)}]{Gardiner1984}
\bibinfo{author}{\bibfnamefont{C.~W.} \bibnamefont{Gardiner}},
  \bibinfo{journal}{Phys. Rev. A} \textbf{\bibinfo{volume}{29}},
  \bibinfo{pages}{2814} (\bibinfo{year}{1984}).

\bibitem[{\citenamefont{Aljunid et~al.}(2009)\citenamefont{Aljunid, Tey, Chng,
  Chen, Lee, Liew, Maslennikov, Scarani, and Kurtsiefer}}]{Kurtsiefer2009}
\bibinfo{author}{\bibfnamefont{S.~A.} \bibnamefont{Aljunid}},
  \bibinfo{author}{\bibfnamefont{M.~K.} \bibnamefont{Tey}},
  \bibinfo{author}{\bibfnamefont{B.}~\bibnamefont{Chng}},
  \bibinfo{author}{\bibfnamefont{Z.}~\bibnamefont{Chen}},
  \bibinfo{author}{\bibfnamefont{J.}~\bibnamefont{Lee}},
  \bibinfo{author}{\bibfnamefont{T.}~\bibnamefont{Liew}},
  \bibinfo{author}{\bibfnamefont{G.}~\bibnamefont{Maslennikov}},
  \bibinfo{author}{\bibfnamefont{V.}~\bibnamefont{Scarani}}, \bibnamefont{and}
  \bibinfo{author}{\bibfnamefont{C.}~\bibnamefont{Kurtsiefer}}, in
  \emph{\bibinfo{booktitle}{2009 Conference on Lasers and Electro-Optics and
  the XIth European Quantum Electronics Conference
  (CLEO\textregistered/Europe-EQEC 2009), Munich, Germany}}
  (\bibinfo{publisher}{IEEE}, \bibinfo{year}{2009}), p.~\bibinfo{pages}{89}.

\bibitem[{\citenamefont{Gradshteyn and Ryzhik}(1994)}]{Gradshteyn1994}
\bibinfo{author}{\bibfnamefont{I.~S.} \bibnamefont{Gradshteyn}}
  \bibnamefont{and} \bibinfo{author}{\bibfnamefont{I.~M.}
  \bibnamefont{Ryzhik}}, \emph{\bibinfo{title}{Table of integrals, series and
  products}} (\bibinfo{publisher}{Academic Press}, \bibinfo{year}{1994}),
  \bibinfo{edition}{5th} ed.

\bibitem[{\citenamefont{Gordon and Ashkin}(1980)}]{Gordon1980}
\bibinfo{author}{\bibfnamefont{J.~P.} \bibnamefont{Gordon}} \bibnamefont{and}
  \bibinfo{author}{\bibfnamefont{A.}~\bibnamefont{Ashkin}},
  \bibinfo{journal}{Phys. Rev. A} \textbf{\bibinfo{volume}{21}},
  \bibinfo{pages}{1606} (\bibinfo{year}{1980}).

\bibitem[{\citenamefont{Cook}(1980)}]{Cook1980}
\bibinfo{author}{\bibfnamefont{R.~J.} \bibnamefont{Cook}},
  \bibinfo{journal}{Phys. Rev. A} \textbf{\bibinfo{volume}{22}},
  \bibinfo{pages}{1078} (\bibinfo{year}{1980}).

\bibitem[{\citenamefont{Cohen-Tannoudji}(1992)}]{CohenTannoudji1992}
\bibinfo{author}{\bibfnamefont{C.}~\bibnamefont{Cohen-Tannoudji}}, in
  \emph{\bibinfo{booktitle}{Fundamental Systems in Quantum Opt., Proceedings of
  the Les Houches Summer School, Session LIII}}, edited by
  \bibinfo{editor}{\bibfnamefont{J.}~\bibnamefont{Dalibard}},
  \bibinfo{editor}{\bibfnamefont{J.}~\bibnamefont{Zinn-Justin}},
  \bibnamefont{and} \bibinfo{editor}{\bibfnamefont{J.~M.}
  \bibnamefont{Raimond}} (\bibinfo{publisher}{North Holland},
  \bibinfo{year}{1992}), pp. \bibinfo{pages}{1--164}.

\bibitem[{\citenamefont{Berg-S{\o}rensen
  et~al.}(1992)\citenamefont{Berg-S{\o}rensen, Castin, Bonderup, and
  M{\o}lmer}}]{BergSorensen1992}
\bibinfo{author}{\bibfnamefont{K.}~\bibnamefont{Berg-S{\o}rensen}},
  \bibinfo{author}{\bibfnamefont{Y.}~\bibnamefont{Castin}},
  \bibinfo{author}{\bibfnamefont{E.}~\bibnamefont{Bonderup}}, \bibnamefont{and}
  \bibinfo{author}{\bibfnamefont{K.}~\bibnamefont{M{\o}lmer}},
  \bibinfo{journal}{J. Phys. B: At. Mol. Opt. Phys.}
  \textbf{\bibinfo{volume}{25}}, \bibinfo{pages}{4195} (\bibinfo{year}{1992}).

\bibitem[{\citenamefont{Horak and Ritsch}(2001)}]{Horak2001}
\bibinfo{author}{\bibfnamefont{P.}~\bibnamefont{Horak}} \bibnamefont{and}
  \bibinfo{author}{\bibfnamefont{H.}~\bibnamefont{Ritsch}},
  \bibinfo{journal}{Phys. Rev. A} \textbf{\bibinfo{volume}{64}},
  \bibinfo{pages}{033422} (\bibinfo{year}{2001}).

\bibitem[{\citenamefont{Rempe et~al.}(1992)\citenamefont{Rempe, Thompson,
  Kimble, and Lalezari}}]{Rempe1992}
\bibinfo{author}{\bibfnamefont{G.}~\bibnamefont{Rempe}},
  \bibinfo{author}{\bibfnamefont{R.~J.} \bibnamefont{Thompson}},
  \bibinfo{author}{\bibfnamefont{H.~J.} \bibnamefont{Kimble}},
  \bibnamefont{and} \bibinfo{author}{\bibfnamefont{R.}~\bibnamefont{Lalezari}},
  \bibinfo{journal}{Opt. Lett.} \textbf{\bibinfo{volume}{17}},
  \bibinfo{pages}{363} (\bibinfo{year}{1992}).

\bibitem[{\citenamefont{Mabuchi and Kimble}(1994)}]{Mabuchi1994}
\bibinfo{author}{\bibfnamefont{H.}~\bibnamefont{Mabuchi}} \bibnamefont{and}
  \bibinfo{author}{\bibfnamefont{H.~J.} \bibnamefont{Kimble}},
  \bibinfo{journal}{Opt. Lett.} \textbf{\bibinfo{volume}{19}},
  \bibinfo{pages}{749} (\bibinfo{year}{1994}).

\end{thebibliography}
\end{document}